\documentclass[11pt]{article}
\usepackage{graphicx} % Required for inserting images
\usepackage{amssymb}
\usepackage{commath}
\usepackage{cancel}
\usepackage{amsmath}
\usepackage{xcolor}
\usepackage[compat=1.0.0]{tikz-feynman}
\usepackage[export]{adjustbox}
\usepackage{subcaption}
\usepackage{slashed}
\usepackage{mathtools}
\usepackage{braket}
\newcommand{\brak}[1]{\left({#1}\right)}
\newcommand{\beqa}{\begin{eqnarray}}
\newcommand{\eeqa}{\end{eqnarray}}
\usepackage[top=2.5cm,bottom=2.5cm,left=1.4cm,right=2.0cm]{geometry}

   \let\b=\beta      \let\d=\delta
 \let\e=\epsilon

 \let\s=\sigma  \let\t=\tau

\def\d{\delta}

\def\e{\epsilon}

\def\b{\beta}

\def\ds#1{#1\kern-1ex\hbox{/}}
\def\dsh{h\kern-1.2ex /}

  \let\D=\Delta   
         \let\S=\Sigma  

\newcommand{\bea}{\begin{eqnarray}}
\newcommand{\eea}{\end{eqnarray}}

\def\nn{\nonumber}
\def\beq{\begin{equation}}
\def\eeq{\end{equation}}

\def\beqn{\begin{eqnarray}}
\def\eeqn{\end{eqnarray}}
\def\ba{\begin{eqnarray}}
\def\ea{\end{eqnarray}}

\begin{document}

\vspace{0.5cm}
\begin{center}
%{\bf \large \\}
{\bf\large The 33311 Left-Right Bilepton Model\\}
\vspace{0.3cm}
\vspace{1cm}
{\bf \large  $^{(1,2,3)}$Claudio Corian\`o, $^{(1)}$Paul H. Frampton and $^{(1,2)}$Dario Melle \\}
\vspace{1cm}
{\it $^{(1)}$Dipartimento di Matematica e Fisica, Universit\`{a} del Salento \\
and \\ 
$^{(2)}$ INFN Sezione di Lecce, Via Arnesano 73100 Lecce, Italy\\
National Center for HPC, Big Data and Quantum Computing\\}
{\it  $^{(3)}$  CNR-Nanotec, Lecce}

\end{center}

\vspace{2.0cm}

\noindent
\begin{abstract}
\noindent
We discuss extension of the electroweak gauge group to $SU(3)_L\times SU(3)_R\times U(1)_L\times U(1)_R$, based on the principle that anomaly cancellation may inherently involve three fermion generations. The explanation for three light quark-lepton families remains, but new particles are introduced which have no TeV scale upper limit on their masses. This approach, originally realized in the framework of the 331 Model, has been consistently defined up to the TeV scale. Our approach maintains the 331 Model's natural explanation for three light quark-lepton families, while introducing new physics with no upper bound on its energy scale.
\end{abstract}
\newpage
\section{Introduction}
A fundamental requirement in any extension of the Standard Model (SM) is the cancellation of gauge anomalies. This principle is crucial because the internal consistency of gauge theories and the renormalisability of their extensions rely on preserving the Ward identities of the quantum action. In the SM, anomaly cancellation is achieved within the electroweak sector on a family-by-family basis, with fermion charge assignments naturally satisfying Diophantine equations that enforce anomaly-free conditions.\\
Despite this success, the SM leaves a central mystery unresolved: the existence of exactly three generations (or families) of quarks and leptons. These families interact directly only through Yukawa couplings, which give rise to the observed mass hierarchies, but the Model does not provide any fundamental explanation for why there are three families. The anomaly cancellation condition, when applied individually to each family, ensures internal consistency but does not inherently account for this replication across three generations.\\
This ambiguity extends to the question of whether, and how, to expand the electroweak sector. While the strong interaction sector, represented by the unbroken \( SU(3)_C \) gauge group of Quantum Chromodynamics (QCD), appears robust and well-understood, the original electroweak gauge group \( SU(2)_L \times U(1)_Y \) offers no explanation for the existence of three sequential fermion families. Consequently, the SM remains incomplete in this regard, motivating the search for extensions that could account for the observed family structure and suggest deeper principles governing the replication of fermion generations.
\noindent
\subsection{Interfamily and the 331 Model} 
The simplest solution to the family replication problem in the Standard Model (SM) involves extending \cite{frampton_chiral_1992,pisano_su3ensuremathbigotimesu1_1992} the electroweak gauge group 
\( SU(2)_L \times U(1)_Y \) to \( SU(3)_L \times U(1)_X \). This extension incorporates the electric charge operator \( Q \) into a more complex structure, defined as

\beq
Q \equiv \frac{1}{2} \lambda_L^3 + \frac{\sqrt{3}}{2} \lambda_L^8 + X \sqrt{\frac{3}{2}} \lambda_L^9,
\eeq
where \( \lambda_L^a \) are the Gell-Mann matrices, normalized such that \( \text{Tr}(\lambda_L^a \lambda_L^b) \equiv \delta^{ab} \). This extension, when combined with the color gauge group \( SU(3)_C \), defines the 331 Model with the gauge symmetry \( SU(3)_C \times SU(3)_L \times U(1)_X \).\\
In this model, each of the three fermion families contains a new quark with charges 
\( Q =( -\frac{4}{3}, -\frac{4}{3}, +\frac{5}{3} )\) respectively for each family. By requiring asymptotic freedom in Quantum Chromodynamics (QCD), which ensures that the strong interaction coupling weakens at high energies, the 331 Model naturally restricts the number of families to three.\\
This is a stronger statement than the observation by Kobayashi and Maskawa \cite{kobayashi_cp-violation_1973} that CP violation emerges naturally with three families, as their framework only implies that there should be \textit{at least} three families, without explaining why there are exactly three. In contrast, the 331 Model directly links the existence of three generations to its gauge structure and anomaly cancellation requirements. \\
The 331 Model predicts a rich spectrum of new particles, including exotic quarks and leptons with unconventional charges and flavor interactions \cite{coriano_su3c_2024}. Among these are the doubly-charged bileptons, which are gauge bosons carrying lepton number and electric charge. One of the most distinctive predictions of the model is the existence of bileptons with charge \( \pm 2 \), which can manifest as same-sign lepton pairs in high-energy collisions. The production of such bileptons provides a clear experimental signature and represents one of the most accessible tests of the 331 Model at TeV-scale energies \cite{corcella_bilepton_2017,corcella_exploring_2018,corcella_non-leptonic_2022,corcella_vector-like_2021,corcella_hunting_2024}.\\
High-energy colliders, such as the Large Hadron Collider (LHC) and its potential upgrades, or future colliders, are well-suited to search for these bileptons. Their discovery would offer compelling evidence for the 331 Model and provide a window into physics beyond the Standard Model.\\
Unlike in the Standard Model, where anomaly cancellation occurs within each family independently, the 331 Model achieves anomaly cancellation across all three families collectively. This \textit{interfamily anomaly cancellation} is a fundamental feature of the model, distinguishing it from other extensions.

\section{33311-Model : chiral fermions and gauge bosons}
\noindent
In the 331-model, the minimal number of beyond the SM fermions
is three, namely one exotically-charged quark in each family and no
additional lepton. For the 33311-model, we can be equally economical
for the quarks but additional leptons are inevitable.

\noindent
Let us define the electric charge $Q$ embedding by
\begin{equation}
Q \equiv \frac{1}{2} \lambda_L^3  + \frac{\sqrt{3}}{2} \lambda_L^8 
+ \frac{1}{2} \lambda_R^3  + \frac{\sqrt{3}}{2} \lambda_R^8 
+ X_L \sqrt{\frac{3}{2}} \lambda_L^9
+ X_R \sqrt{\frac{3}{2}} \lambda_R^9
\label{Q33311}
\end{equation}

\noindent
For the first family of quarks, in a notation
$(3_C, 3_L, 3_R)_{X_L,X_R}$ we include
\begin{equation}
(3, 3, 1)_{-1/3,0} + (3, 1, 3^*)_{0,+1/3}
\label{quarks1}
\end{equation}
which correspond to multiplets
\begin{equation}
\left( \begin{array}{c} u \\d\\D  
\end{array} \right) +
\left( \begin{array}{c} \bar{D} \\ \bar{d} \\ \bar{u}  
\end{array} \right)
\end{equation}
The second family quarks are in the representations of Eq.(\ref{quarks1}),
for the multiplets
\begin{equation}
\left( \begin{array}{c} c \\s\\S  
\end{array} \right) +
\left( \begin{array}{c} \bar{S} \\ \bar{s} \\ \bar{c}  
\end{array} \right)
\end{equation}

\bigskip

\noindent
For the third family we replace Eq.(\ref{quarks1})  by
\begin{equation}
(3, 3^*, 1)_{+2/3,0} + (3, 1, 3)_{0,-2/3}
\label{quarks3}
\end{equation}
and assign quarks to the multiplets
\begin{equation}
\left( \begin{array}{c} T \\t\\b  
\end{array} \right) +
\left( \begin{array}{c} \bar{b} \\ \bar{t} \\ \bar{T}  
\end{array} \right)
\end{equation}
for every of this field the corresponding right handed singlets are added with the following quantum numbers
\begin{align}
	&\begin{matrix}
		(u^c)_L\\
		(c^c)_L\\ 
		(t^c)_L
	\end{matrix} \rightarrow \brak{{{3}}^*, {1}, {1}}_{-\frac{2}{3},0}\ ,
	&&\begin{matrix}
	(\bar u^c)_L\\
	(\bar c^c)_L\\ 
	(\bar t^c)_L
	\end{matrix} \rightarrow \brak{{ {3}}^*, {1}, {1}}_{0,\frac{2}{3}}\ ,\\\nn\\
	&\begin{matrix}
		(d^c)_L\\
		(s^c)_L\\
		(b^c)_L
	\end{matrix} \rightarrow \brak{{ {3}}^*, {1}, {1}}_{\frac{1}{3},0}\ ,
	&&\begin{matrix}
		({\bar{ d}}^c)_L\\
		(\bar s^c)_L\\
		({\bar b } ^c)_L
	\end{matrix} \rightarrow \brak{{ {3}}^*, {1}, {1}}_{0,-\frac{1}{3}}\ ,\\\nn\\
	&\begin{matrix}
		(D^c)_L\\(S^c)_L
	\end{matrix} \rightarrow \brak{{ {3}}^*, {1}, {1}}_{\frac{4}{3},0}\ ,
	&&\begin{matrix}
		(\bar D^c)_L\\(\bar S^c)_L
	\end{matrix} \rightarrow \brak{{ {3}}^*, {1}, {1}}_{0,-\frac{4}{3}}\ ,\\\nn\\
		&\begin{matrix}
		(T^c)_L
	\end{matrix} \rightarrow \brak{{ {3}}^*, {1}, {1}}_{-\frac{5}{3},0}\ ,
	&&\begin{matrix}
		(\bar T^c)_L
	\end{matrix} \rightarrow \brak{{ {3}}^*, {1}, {1}}_{0,\frac{5}{3}}\ ,
\end{align}

\bigskip

\noindent
Similarly to the  331-model, the leptons are neutral with respect
to $U(1)_{LX} \times U(1)_{RX}$ and assigned to
\begin{equation}
3[ (1, 3^*, 1)_{0,0} + (1, 1, 3^*)_{0.0}]
\label{leptons33311}
\end{equation}
\bigskip

\noindent
The lepton multiplets are
\begin{equation}
\left( \begin{array}{c} e^+ \\\nu_e\\e^- 
\end{array} \right) +
\left( \begin{array}{c} E^+\\ N_e\\ E^-
\end{array} \right) ~~~
\left( \begin{array}{c} \mu^+ \\ \nu_{\mu}\\ \mu^- 
\end{array} \right) +
\left( \begin{array}{c} M^+\\  N_{\mu}\\ M^-
\end{array} \right) ~~~\left( \begin{array}{c} \tau^+ \\ \nu_{\tau} \\ \tau^-
\end{array} \right) +
\left( \begin{array}{c} {\cal T}^+ \\ N_{\tau} \\ {\cal T}^-
\end{array} \right)
\end{equation}
and for each family we need to include an additional
charged lepton and neutrino.  The three families
are treated democratically.

\noindent
While the strong interaction QCD  remains with its eight gluons,
the electroweak sector has expanded its number of gauge bosons
from four is the standard model to nine in the 331-model to
eighteen in the 33311-model.  We denote by an $R$ subscript
the new gauge bosons as follows
\begin{equation}
(W^{\pm}, Z, \gamma, Y^{++},Y^+, Y^-,Y^{--}, Z^{'};
W_R^{\pm}, Z_R, \gamma_R, Y^{++}_R,Y^+_R, Y^-_R,Y^{--}_R, Z^{'}_R)
\label{gauge}
\end{equation}

\section{Anomaly cancellation in 33311}

There are thirteen potentially troublesome triangle anomalies
which are, in an obvious notation,
\begin{equation}
SU(3)_C^3, \qquad SU(3)_C^2 U(1)_{XL}, \qquad SU(3)_L^3, \qquad  SU(3)_L^2 U(1)_{XL}, \qquad U(1)_{XL}^3
\label{anomalies1}
\end{equation}
as well as
\begin{eqnarray}
SU(3)_C^2U(1)_{XR} \qquad SU(3)_L^2U(1)_{XR}\qquad  SU(3)_R^3 \qquad  SU(3)_R^2U(1)_{XR},\nonumber  \\
SU(3)_R^2U(1)_{XR}\qquad U(1)_{XL}^2 U(1)_{XR}\qquad U(1)_{XL}U(1)_{XR}^2 \qquad U(1)_{XR}^3
\label{anomalies2}
\end{eqnarray}
\noindent
The first five in Eq.(\ref{anomalies1}) cancel exactly as in the
331-model, provided that there exist exactly three families. We note
that all higher multiples of three are excluded by the requirement
of QCD asymptotic freedom.
\noindent
We must therefore examine the anomalies in Eq.(\ref{anomalies2}). That these all cancel
follows from\\\\
\noindent
{\underline{$SU(3)_C^2U(1)_{XR}$.}\\
Six colour triplets with $X_R=+1/3$\\
Three colour triplets with $X_R=-2/3$.\\\\}
\noindent 
\underline{$SU(3)_L^2U(1)_{XR}$.}\\
Two triplets with $X_R=-1/3.$\\
One triplet with $X_R =- 2/3.$\\
One triplet with $X_R=0$\\\\
\noindent
\underline{$SU(3)_R^3$.}\\
Six triplets.\\
Six antitriplets.\\\\
\noindent\\
\underline{$SU(3)_R^2U(1)_{XL}$. }\\
Two triplets with $X_L=-1/3$.\\
One antriplet wuth $X=+2/3$.
One triplet with $X_R=0$.\\\\
\noindent
\underline{$SU(3)_R^2U(1)_{XR}$.}\\
Two antitriplets with $X_R=+1/3$.\\
One triplet with $X_R=-2/3$.\\
One.triplet with $X_R =0$.\\\\
\noindent
\underline{$U(1)_{XL}^2 U(1)_{XR}$.}\\
No states with $X_L \neq 0$ and $X_R \neq 0$.\\\\
\noindent
\underline{$U(1)_{XL} U(1)_{XR}^2$}.\\
No states with $X_L \neq 0$ and $X_R \neq 0$.\\\\
\noindent
\underline{$U(1)_{XR}^3$.}\\
Eighteen states have $X_R =1/3$.\\
Nine states have $X_R = -2/3$.

\section{Scalar representations and stages of symmetry breaking}
In the context of the 331 Model, the scalar sector is responsible for the spontaneous breaking of the gauge symmetry at different energy scales. The scalar fields are organized into representations under the extended gauge symmetry, and their vacuum expectation values (VEVs) trigger symmetry breaking, ultimately reducing the gauge group down to the Standard Model at low energies. Here is an explanation of how the scalar sector works in terms of representations.\\
\noindent
The 33311-model has a minimum of three energy scale
associated with its spontaneous symmetry breaking.\
 The lowest one is the weak scale $M_W \simeq 248$ GeV.\\
The middle one is the 331 scale, $M_{331}\sim 4$ TeV \cite{martinez_landau_2007}.
Unlike the middle scale, the highest one has no upper bound $M_{331} < M_{33311}<\infty$.\\
In the asymptotic limit $M_{33311} \rightarrow \infty$,  the 33311-model reduces to the
331-model.
\noindent
Thus, although the 33311-Model shares with the 331-Model the explanation
for the existence of three families, it does not share the attractive property
that all the new physics is constrained to be in the TeV region.
\noindent
In a minimal scenario, the spontaneous symmetry breaking can occur 
at three distinct energy scales.\\
 Using again the notation \( (3_C, 3_L, 3_R)_{X_L, X_R} \), we here list a set of scalar fields whose VEVs can be sufficient to spontaneously break the symmetry, as indicated in the previous section.

\noindent
\subsection{Breaking at the Highest Scale \( M_{33311} \)}
At the highest scale \( M_{33311} \), the gauge group \( SU(3)_R \times U(1)_{X_L} \times U(1)_{X_R} \) is broken down. The following scalar fields are introduced with VEVs that trigger this breaking:

\begin{itemize}
    \item \(\Phi(1, 1, 1)_{+1, -1}\): This scalar field is a singlet under \( SU(3)_L \) and \( SU(3)_R \) but carries charges under \( U(1)_{X_L} \) and \( U(1)_{X_R} \). Its VEV breaks the \( U(1)_{X_L} \times U(1)_{X_R} \) symmetry down to a single \( U(1)_X \).
    
    \item \(\Phi(1, 1, 3)_{0,0}\): This is a triplet under \( SU(3)_R \) and is neutral under \( SU(3)_L \) and \( U(1)_X \). Its VEV breaks \( SU(3)_R \) completely, leaving only the unbroken gauge group \( SU(3)_L \times U(1)_X \). The model requires two of these scalar triplets $3_R$  misaligned. 
\end{itemize}

\noindent
At this scale 

\beq
U(1)_{X_L} \times U(1)_{X_R} \rightarrow U(1)_X
\eeq

\beq
SU(3)_R \rightarrow \text{nothing}
\eeq

\subsection{At scale \( M_{331} \)}

In the 33311 Model, the scalar sector at the intermediate scale \( M_{331} \) is identical to the scalar sector of the 331 Model \cite{tully_scalar_2003,costantini_theoretical_2020}. The scalar field listed here plays a specific role in this symmetry-breaking process, ensuring that the correct gauge bosons acquire masses while preserving the Standard Model structure at lower energies.

\begin{itemize}
    \item \(\Phi((1, 3, 1)_{+1,0}\): 
    This field is a triplet under \( SU(3)_L \) and a singlet under \( SU(3)_R \), carrying a charge under \( U(1)_X \).
    Its primary function is to break the \( SU(3)_L \) symmetry down to \( SU(2)_L \), the gauge group of the Standard Model electroweak interactions.
\end{itemize}

\noindent
This field contributes to the mass generation of the extended gauge bosons which are also present in the 331 Model, maintains anomaly cancellation, and aligns the vacuum expectation values to ensure that the electroweak \( SU(2)_L \times U(1)_Y \) structure remains intact:
\beq
SU(3)_L \times U(1)_X \rightarrow SU(2)_L \times U(1)_Y
\eeq

\subsection{At scale \( M_W \)}
At the electroweak scale \( M_W \), the scalar fields are 
\begin{itemize}
    \item \(\Phi(1, 3, 1)_{0,0}\): 
   This is another triplet under \( SU(3)_L \) and a singlet under \( SU(3)_R \), but it is neutral under \( U(1)_X \).
   
   \item \(\Phi(1, 3, 1)_{-1,0}\): 
   This scalar is similar to \(\Phi(1, 3, 1)_{0,0}\) in terms of its \( SU(3)_L \) triplet structure, but it carries a charge of \(-1\) under \( U(1)_X \).
   
   \item \(\Phi(1, 6, 1)_{0,0}\): 
   This field is a symmetric representation under \( SU(3)_L \), often written as a sextet, and is neutral under both \( SU(3)_R \) and \( U(1)_X \).
\end{itemize}
which are necessary to break the symmetry to the electomagnetic $U(1)$
\beq
 SU(2)_L \times U(1)_Y  \to  U(1)_Q .
\eeq

\section{Discussion}
\noindent
In traditional GUTs, like $SU(5)$, all three families of fermions are treated as identical, sequential generations under a single unified gauge group\cite{georgi_unity_1974,frampton_cosmological_2002}. However, the 33311 Model breaks away from this assumption by requiring a non-sequential arrangement of families to cancel anomalies. This structure fundamentally differentiates the 33311 Model from typical GUTs in several ways.
This distinction becomes crucial above the TeV scale, where the 33311 Model's non-sequential family organization ensures anomaly cancellation and theoretical consistency.\\
 Traditional GUTs predict proton decay as an outcome of unifying quarks and leptons within a single representation. However, the 33311 Model sidesteps this issue by not embedding quarks and leptons in the same way. Thus, it provides an explanation for the observed stability of the proton without conflicting with experimental bounds on proton decay.

\providecommand{\href}[2]{#2}\begingroup\raggedright\endgroup

%\bibliographystyle{jhep}
%\bibliography{bib}
\end{document}